\documentclass{article}
\usepackage{spconf,amsmath,graphicx}
\usepackage{amsfonts}
\usepackage{booktabs}
\usepackage{multirow}
\usepackage[hidelinks]{hyperref}
\usepackage{siunitx}
\usepackage{mathtools}
\usepackage[subrefformat=parens,aboveskip=1pt,belowskip=7pt]{subcaption}
\usepackage{comment}
\usepackage[linesnumbered,ruled,vlined]{algorithm2e}
\usepackage{color}

\def\equationautorefname~#1\null{(#1\null)}

\let\oldnl\nl
\newcommand{\nonl}{\renewcommand{\nl}{\let\nl\oldnl}}
\newlength\mylen

\SetCommentSty{mycommfont}

\newcommand{\ctext}[1]{\raise0.2ex\hbox{\textcircled{\scriptsize{#1}}}}

\usepackage{xspace}
\makeatletter
\DeclareRobustCommand\onedot{\futurelet\@let@token\@onedot}
\def\@onedot{\ifx\@let@token.\else.\null\fi\xspace}
\def\eg{\emph{e.g}\onedot} 
\def\ie{\emph{i.e}\onedot}

\makeatother

\newcommand{\gray}[1]{\textcolor[gray]{0.55}{#1}}

\newcommand{\dersingle}{$\text{DER}_\text{1ch}$\xspace}
\newcommand{\dermulti}{$\text{DER}_\text{4ch}$\xspace}

\def\equationautorefname~#1\null{(#1\null)}
\renewcommand{\subfigureautorefname}{\figureautorefname}

\renewcommand{\sectionautorefname}{Section}
\renewcommand{\subsectionautorefname}{\sectionautorefname}

\let\orgautoref\autoref

\renewcommand{\autoref}[1]
{%
\def\figureautorefname{Fig.}%
\def\subfigureautorefname{Fig.}%
\def\sectionautorefname{Sec.}%
\def\subsectionautorefname{Sec.}%
\def\subsectionautorefname{Sec.}%
\orgautoref{#1}%
}

\newcommand{\vect}[1]{\mbox{\boldmath $#1$}}

\newcommand{\norm}[1]{\left\lVert#1\right\rVert}

\newcommand{\trans}[1]{#1^\mathsf{T}}

\def\appendixautorefname~#1\null{~#1 \null}
\newcommand{\relmid}{\mathrel{}\middle|\mathrel{}}

\makeatletter
\newcommand{\figcaption}[1]{\def\@captype{figure}\caption{#1}}
\newcommand{\tblcaption}[1]{\def\@captype{table}\caption{#1}}
\makeatother

\def\tablescale{0.91}

\title{Mutual learning of single- and multi-channel\\end-to-end neural diarization}

\name{Shota Horiguchi$^1$ \quad Yuki Takashima$^1$ \quad Shinji Watanabe$^2$ \quad Paola Garc\'{i}a$^3$}
\address{
  $^1$Hitachi, Ltd., Japan\\
  $^2$Carnegie Mellon University, USA\\
  $^3$Johns Hopkins University, USA}
\begin{document}
\ninept
\abovedisplayskip=4pt
\belowdisplayskip=4pt

\setlength\textfloatsep{10pt}
\setlength\abovecaptionskip{4pt}
\setlength\belowcaptionskip{4pt}

\maketitle
\begin{abstract}
Due to the high performance of multi-channel speech processing, we can use the outputs from a multi-channel model as teacher labels when training a single-channel model with knowledge distillation.
To the contrary, it is also known that single-channel speech data can benefit multi-channel models by mixing it with multi-channel speech data during training or by using it for model pretraining.
This paper focuses on speaker diarization and proposes to conduct the above bi-directional knowledge transfer alternately.
We first introduce an end-to-end neural diarization model that can handle both single- and multi-channel inputs.
Using this model, we alternately conduct i) knowledge distillation from a multi-channel model to a single-channel model and ii) finetuning from the distilled single-channel model to a multi-channel model.
Experimental results on two-speaker data show that the proposed method mutually improved single- and multi-channel speaker diarization performances.
\end{abstract}
\begin{keywords}
Speaker diarization, EEND, multi-channel, knowledge distillation, transfer learning, mutual learning
\end{keywords}
\section{Introduction}
\label{sec:intro}
Speech processing under noisy and reverberant environments or the existence of multiple speakers expands the practicality of speech applications.
While single-channel solutions for such conditions are widely studied, multi-channel approaches have shown promising performance in various speech applications such as speech recognition \cite{sainath2017multichannel,ochiai2017multichannel}, speech separation \cite{wang2018multi,luo2020endtoend}, speaker recognition \cite{taherian2019deep}, and speaker diarization \cite{kinoshita2020tackling,medennikov2020targetspeaker_short,horiguchi2022multichannel}.
Especially, multi-channel processing based on distributed microphones rather than microphone-array devices is attracting much attention for its high versatility \cite{horiguchi2022multichannel,araki2017meeting,luo2019fasnet,yoshioka2022vararray}.

Since multi-channel speech processing is powerful, its outputs are sometimes used as teacher labels when training a single-channel model, which is known as knowledge distillation or teacher-student learning \cite{bucilua2006model,hinton2014distilling}.
On the other hand, it has been reported that single-channel data is still useful in training multi-channel models, \eg, single-channel pretraining \cite{heymann2017beamnet,zhang2020endtoend,an2022exploiting} and simultaneous use of single- and multi-channel data \cite{chang2019mimospeech,horiguchi2022multichannel,an2022exploiting}.
This can be because the information captured by single- and multi-channel models are different.
For example, when considering speech separation or speaker diarization, single-channel methods must rely on speaker characteristics, while multi-channel methods can use spatial information additionally (or even only).
Another study demonstrated that incorporating spectral and spatial information boosts speech separation performance \cite{wang2018spatial}.
Let us consider a multi-channel model that can also handle single-channel inputs.
Using single-channel data to train such a multi-channel model avoids falling into local minima that rely too heavily on spatial information and allow the model to benefit more from speaker characteristics \cite{horiguchi2022multichannel}.
Here a research question arises---does iterative knowledge distillation from multi-channel to single-channel model and finetuning from single-channel to multi-channel model improve the performance of both single and multi-channel speech processing?

\begin{figure}[t]
    \centering
    \includegraphics[width=\linewidth]{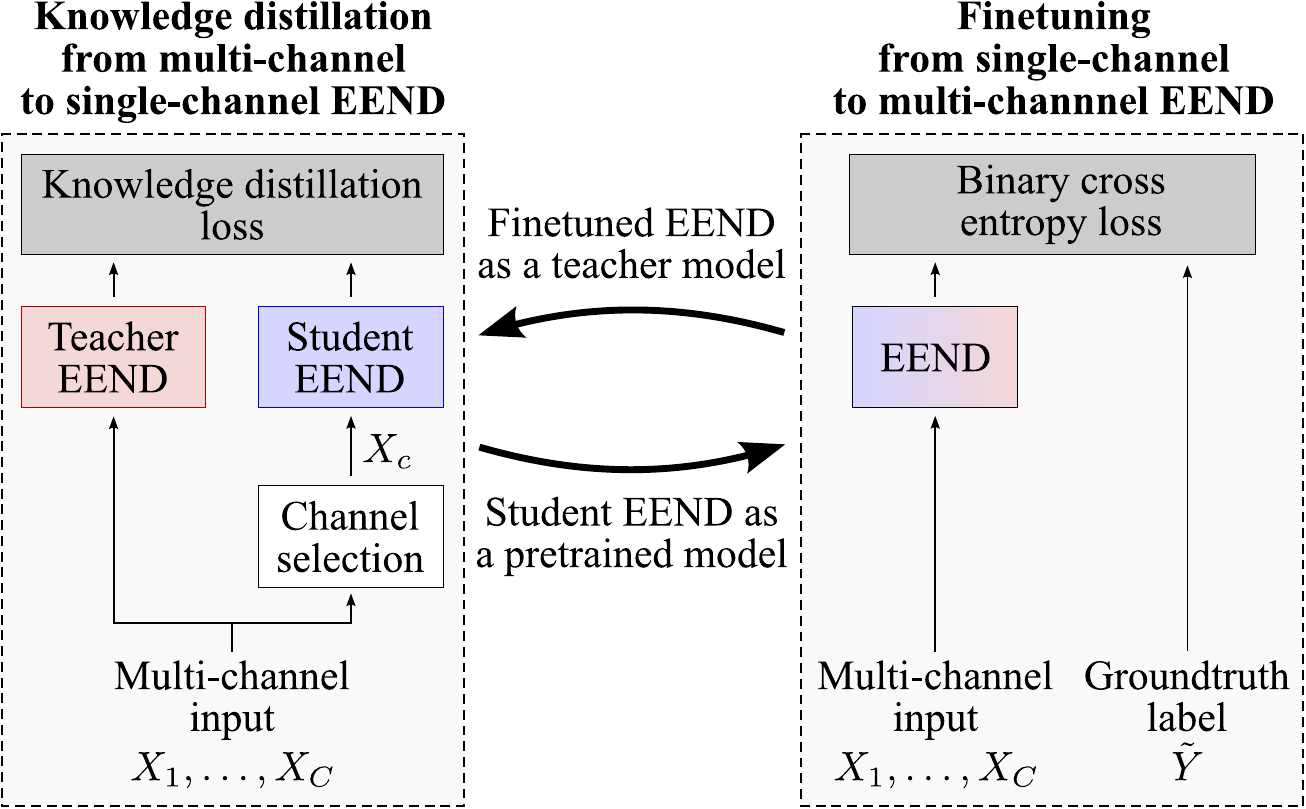}
    \caption{Mutual learning of single- and multi-channel EEND.}
    \label{fig:overview}
\end{figure}

Given that question as motivation, this paper proposes a mutual learning method of single- and multi-channel end-to-end neural diarization (EEND), illustrated in \autoref{fig:overview}.
We focus specifically on speaker diarization here, but the method can be applied to other speech processing tasks such as speech recognition and separation.
We first introduce a co-attention-based multi-channel EEND model invariant to the number and geometry of microphones.
The multi-channel model is designed to be identical to the conventional Transformer-based single-channel EEND given single-channel inputs.
We conduct the following processes iteratively: i) distilling the knowledge from multi-channel EEND to single-channel EEND (\autoref{fig:overview} left) and ii) finetuning from the distilled single-channel EEND to multi-channel EEND (\autoref{fig:overview} right)\footnote{The proposed method can also be applied to two multi-channel models, $u$-channel and $v$-channel models with $u<v$.}.
We demonstrate that the proposed method mutually improves both single- and multi-channel speaker diarization performance.

\section{Related work}
\subsection{Speaker diarization}
Speaker diarization is a task to determine who is speaking when from input audio.
It has long been conducted by clustering speaker embeddings extracted from speech segments \cite{anguera2012speaker}, but recently end-to-end methods have attracted much attention, such as EEND \cite{horiguchi2022encoderdecoder,horiguchi2022multichannel}, target-speaker voice activity detection \cite{medennikov2020targetspeaker_short}, and recurrent selective hearing networks \cite{kinoshita2020tackling}.
One reason is that optimization is simple because everything is completed in one model.
Since the one-model approach also makes it possible to apply knowledge transfer techniques such as knowledge distillation and finetuning to the entire network easily, this paper focuses on end-to-end methods, especially EEND.

\subsection{Knowledge distillation in speech applications}
Knowledge distillation or teacher-student learning is a scheme to train a student model to mimic a well-trained teacher model \cite{bucilua2006model,hinton2014distilling}.
It is widely used in speech applications such as speech recognition \cite{li2014learning} and separation \cite{chen2018distilled}.
One typical use case is knowledge distillation between different network architectures: a large model to a small model \cite{li2014learning,tan2018knowledge}, a normal model to a binarized model \cite{chen2018distilled}, an ensemble of models to a single model \cite{fukuda2017efficient}, and a high-latency model to a streaming model \cite{kim2018improved}.

The other type of knowledge distillation, which we focus on in this paper, is based on different inputs, while the network architectures are not necessarily different.
In some studies on unsupervised domain adaptation of speech recognition \cite{meng2019domain} and speaker verification \cite{zhang2019fully}, a far-field model is trained with knowledge distillation by using a close-talk model as a teacher, both of which take single-channel signals as inputs.
Another series of studies leverage multi-channel signals; a student model is trained so that the output when the noisy features are input is close to the output of the teacher model when the enhanced features are input \cite{watanabe2017studentteacher,subramanian2018student}.
Here, the enhanced features are calculated from a multi-channel signal using beamforming and the input to the model is still single-channel; thus, it is not applicable to speaker diarization where there is more than one speaker to be enhanced.
In the context of continuous speech separation \cite{chen2020continuous}, a student VarArray model \cite{yoshioka2022vararray} is trained to produce similar outputs to a teacher model with a fewer number of channels \cite{wang2022leveraging}.
This paper, in contrast, tackles multi- to single-channel knowledge distillation with an end-to-end model rather than multi- to multi-channel knowledge distillation as in \cite{wang2022leveraging}.

\section{Formulation of Single-Channel End-to-End Neural Diarization}
To facilitate the explanation of the proposed method, in this section we first formulate single-channel EEND and Transformer encoders contained in it.
For simplicity, we omit the bias parameters of each fully-connected layer from the formulation.

\subsection{Overview}\label{sec:overview}
As a single-channel model, we used attractor-based EEND (EEND-EDA) \cite{horiguchi2022encoderdecoder}, in which speaker-wise speech activities are calculated from speaker-wise attractors and frame-wise embeddings.
Input $T$-length $F$-dimensional frame-wise acoustic features $X\in\mathbb{R}^{F\times T}$ are first converted using a position-wise fully-connected layer parameterized by $W_0\in\mathbb{R}^{D\times F}$ and layer normalization as 
\begin{equation}
    E^{(0)}=\mathsf{LayerNorm}\left(W_0 X\right)\in\mathbb{R}^{D\times T}.\label{eq:change_dimension}
\end{equation}
The resulting frame-wise embeddings are further converted using $N$-stacked Transformer encoders \cite{vaswani2017attention} without positional encodings.
The transition in $n$-th encoder layer for $1\leq n\leq N$ is denoted as
\begin{equation}
    E^{(n)}=\mathsf{TransformerEncoder}\left(E^{(n-1)}\right)\in\mathbb{R}^{D\times T}.\label{eq:encoder}
\end{equation}
Then, $S$ speakers' speech activities $Y$ are estimated based on inner products between the frame-wise embeddings from the last encoder $E^{(N)}$ and speaker-wise attractors $B$ as
\begin{align}
    B&=\mathsf{EDA}\left(E^{(N)}\right)\in\left(-1,1\right)^{D\times S},\label{eq:attractors}\\
    Z&=\left(\trans{B}E^{(N)}\right)\in\mathbb{R}^{S\times T},\label{eq:logits}\\
    Y&=\sigma\left(Z\right)\in\left(0,1\right)^{S\times T}\label{eq:posteriors}
\end{align}
where $\mathsf{EDA}$ is an encoder-decoder-based attractor calculation module, $\trans{\left(\cdot\right)}$ denotes matrix transpose, and $\sigma\left(\cdot\right)$ is an element-wise sigmoid operation.
Note that the inner products $Z$ between the embeddings and attractors become logits of the speech activities $Y$.

The speech activities are optimized to minimize the permutation-free loss, which is defined as
\begin{equation}
    \mathcal{L}_{\text{BCE}}\left(\vect{\Theta}\relmid X,\tilde{Y}\right)=\frac{1}{TS}\min_P\mathsf{BCE}\left(\tilde{Y}, PY\right)\label{eq:bce_loss}
\end{equation}
where $P\in\left\{0,1\right\}^{S\times S}$ denotes a $S\times S$ permutation matrix , $\tilde{Y}\in\left\{0,1\right\}^{S\times T}$ is the groundtruth speech activities, and $\mathsf{BCE}\left(\cdot,\cdot\right)$ is the summation of the element-wise binary cross entropy.
$\vect{\Theta}$ is a set of parameters of the network.

\subsection{Detailed formulation of Transformer encoder}\label{sec:transformer_encoder}
\begin{figure}[t]
    \centering
    \subfloat[][Transformer encoder]{\includegraphics[width=\linewidth]{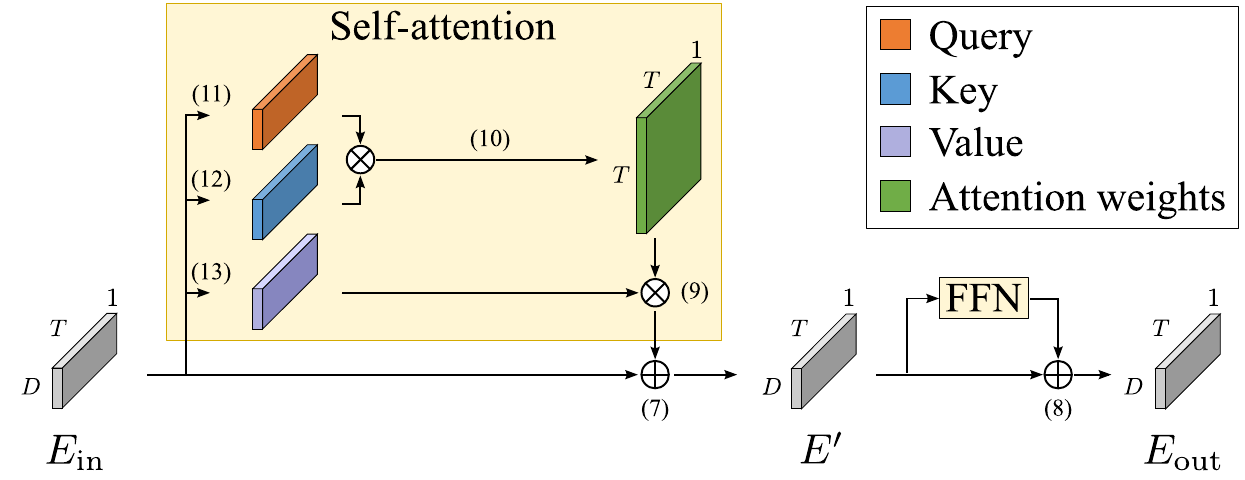}\label{fig:transformer}}\\
    \subfloat[][Co-attention encoder]{\includegraphics[width=\linewidth]{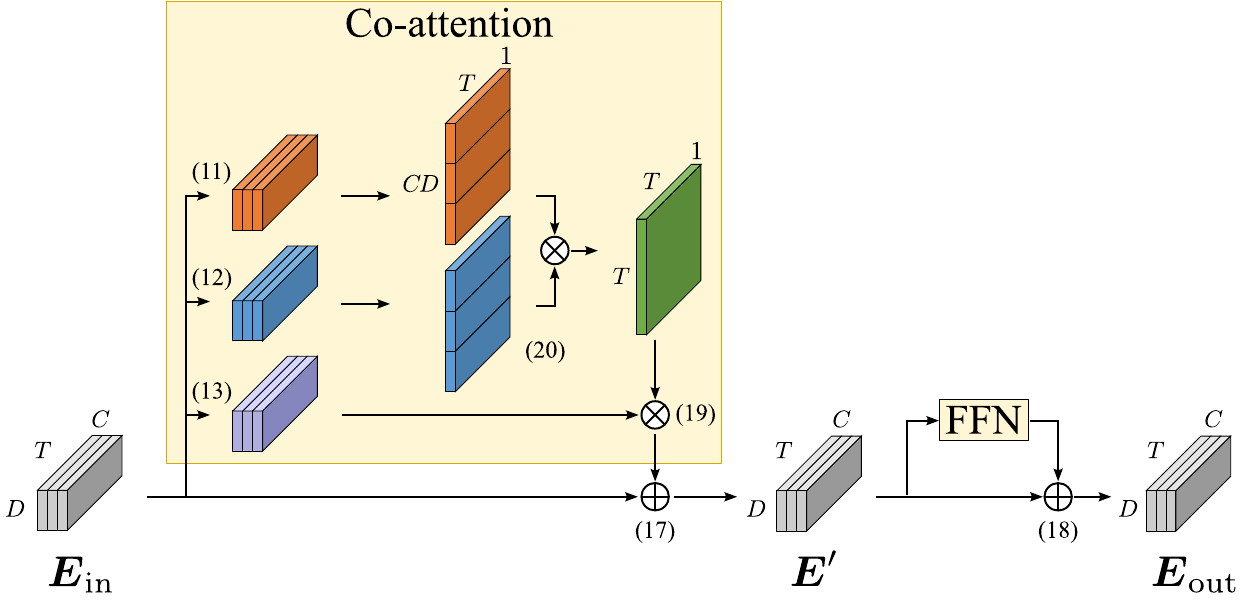}\label{fig:coattention}}
    \caption{The architectures of encoders used in this paper. $D$: the dimensionality of embeddings, $T$: sequence length, $C$: the number of channels.}
    \label{fig:encoder_blocks}
\end{figure}

Given an input embedding sequence $E_\text{in}\in\mathbb{R}^{D\times T}$, each Transformer encoder layer in \autoref{eq:encoder} converts it into $E_\text{out}\in\mathbb{R}^{D\times T}$ as follows: 
\begin{align}
    E'&=\mathsf{LayerNorm}\left(E_\text{in}+\mathsf{MA}\left(E_\text{in},E_\text{in},E_\text{in}\right);\vect{\Phi}\right)\in\mathbb{R}^{D\times T},\\
    E_\text{out}&=\mathsf{LayerNorm}\left(E'+\mathsf{FFN}\left(E';\vect{\Psi}\right)\right),
\end{align}
where $\vect{\Phi}$ and $\vect{\Psi}$ are sets of parameters, and $\mathsf{MA}$ and $\mathsf{FFN}$ denote multi-head scaled dot-product attention and a feed-forward network that consists of two fully-connected layers, respectively.
Note that we omit the layer index $(n)$ for simplicity.
The processes above are illustrated in \autoref{fig:transformer}.

The multi-head scaled dot-product attention $\mathsf{MA}$ is calculated from query $Q\in\mathbb{R}^{d_k\times T}$, key $K\in\mathbb{R}^{d_k\times T}$, and value $V\in\mathbb{R}^{d_v\times T}$ as follows:
\begin{equation}
    \mathsf{MA}\left(Q,K,V;\vect{\Phi}\right)=
    W_O\begin{bmatrix}V^{(1)}A^{(1)\mathsf{T}}\\[-5pt]\vdots\\[-1pt]V^{(h)}A^{(h)\mathsf{T}}\end{bmatrix}\in\mathbb{R}^{d_v\times T},\label{eq:multihead_attention}
\end{equation}
\begin{align}
    A^{(i)}&=\mathsf{softmax}\left(\frac{Q^{(i)\mathsf{T}}K^{(i)}}{\sqrt{d_k/h}}\right)\in\left(0,1\right)^{T\times T},\label{eq:attention}\\
    Q^{(i)}&=W_Q^{(i)}Q\in\mathbb{R}^{\frac{d_k}{h}\times T},\label{eq:q}\\
    K^{(i)}&=W_K^{(i)}K\in\mathbb{R}^{\frac{d_k}{h}\times T},\label{eq:k}\\
    V^{(i)}&=W_V^{(i)}V\in\mathbb{R}^{\frac{d_v}{h}\times T},\label{eq:v}
\end{align}
where $h$ is the number of heads, each of which is indexed by $i$.
$\mathsf{softmax}\left(\cdot\right)$ is a column-wise softmax operation.
$W_Q^{(i)}$, $W_K^{(i)}$, $W_V^{(i)}$, and $W_O$ are weight parameters.
The set of parameters $\vect{\Phi}$ is defined as
\begin{equation}
    \vect{\Phi}\coloneqq\left\{W_O\right\}\cup\bigcup_{1\leq i\leq h}\left\{W_Q^{(i)},W_K^{(i)},W_V^{(i)}\right\}.
\end{equation}

\section{Proposed method}
This study aims to improve both single- and multi-channel diarization by alternating between knowledge distillation from a multi-channel model to a single-channel model and finetuning from a single-channel model to a multi-channel model (\autoref{fig:overview}).
To achieve this, we first introduce the EEND model that can handle both multi-channel and single-channel inputs with exactly the same sets of parameters $\vect{\Phi}$ and $\vect{\Psi}$ in \autoref{sec:model}.
Then, we describe a mutual learning method of single- and multi-channel models in \autoref{sec:mutual_learning}.

\subsection{Multi-channel EEND}\label{sec:model}
For multi-channel diarization, we used a co-attention-based extension of EEND, which was originally proposed in \cite{horiguchi2022multichannel}.
In this paper, we used the simplified version of the encoder proposed in \cite{horiguchi2022multichannel}, which has the same number of parameters as the Transformer encoder introduced in \autoref{sec:transformer_encoder}.

Given frame-wise acoustic features for each of $C$ channels $\vect{X}\coloneqq\left[X_1,\dots,X_C\right]\in\mathbb{R}^{F\times T\times C}$, each is first converted using \autoref{eq:change_dimension} to obtain frame-wise embeddings for each channel independently $\left[E_1^{(0)},\dots,E_c^{(0)}\right]\eqqcolon\vect{E}^{(0)}$.
Then, the resulting tensor is processed using $N$-stacked co-attention encoder.
The $n$-th encoder layer converts $\vect{E}^{(n-1)}\in\mathbb{R}^{D\times T\times C}$ into a tensor of the same shape $\vect{E}^{(n)}\in\mathbb{R}^{D\times T\times C}$ as follows:
\begin{equation}
    \vect{E}^{(n)}=\mathsf{CoAttentionEncoder}\left(\vect{E}^{(n-1)}\right).\label{eq:coattentionencoder}
\end{equation}
The output from the last encoder $\vect{E}^{(N)}\coloneqq\left[E_1^{(N)}\dots,E_C^{(N)}\right]$ is averaged across channels as
\begin{equation}
    E^{(N)}=\frac{1}{C}\sum_{c=1}^C E_c^{(N)},\label{eq:average_channel}
\end{equation}
and it is used for calculation of speech activities with \autoref{eq:attractors}--\autoref{eq:posteriors}.

The co-attention encoder layer in \autoref{eq:coattentionencoder} was illustrated in \autoref{fig:coattention}. It has the same set of parameters to a Transformer encoder and converts input $\vect{E}_\text{in}=\left[E_{\text{in},c}\right]_{c=1}^C\in\mathbb{R}^{D\times T\times C}$ into $\vect{E}_\text{out}=\left[E_{\text{in},c}\right]_{c=1}^C\in\mathbb{R}^{D\times T\times C}$ as
\begin{align}
    E'_c&=\mathsf{LayerNorm}\left(E_{\text{in},c}+\mathsf{MCA}\left(\vect{E}_\text{in},\vect{E}_\text{in},E_{\text{in},c};\vect{\Phi}\right)\right),\label{eq:multi_mca}\\
    E_{\text{out},\text{c}}&=\mathsf{LayerNorm}\left(E'_c+\mathsf{FFN}\left(E'_c;\vect{\Psi}\right)\right).\label{eq:multi_ffn}
\end{align}
Here, $\mathsf{MCA}$ is the multi-head scaled dot-product co-attention.
Given multi-channel query $\vect{Q}=\left[Q_c\right]_{c=1}^C\in\mathbb{R}^{d_k\times T\times C}$, key $\vect{K}=\left[K_c\right]_{c=1}^C\in\mathbb{R}^{d_k\times T\times C}$, and value $\vect{V}=\left[V_c\right]_{c=1}^C\in\mathbb{R}^{d_v\times T\times C}$, multi-head co-attention is defined as
\begin{equation}
    \mathsf{MCA}\left(\vect{Q},\vect{K},V_c;\vect{\Phi}\right)=
    W_O\begin{bmatrix}V_c^{(1)}A^{(1)\mathsf{T}}\\[-5pt]\vdots\\[-1pt]V_c^{(h)}A^{(h)\mathsf{T}}\end{bmatrix}\in\mathbb{R}^{d_v\times T},\label{eq:multihead_coattention}
    \end{equation}
\begin{align}
    A^{(i)}&=\mathsf{softmax}\left(\frac{\left[Q^{(i)\mathsf{T}}_1,\dots,Q^{(i)\mathsf{T}}_C\right]\trans{\left[K^{(i)\mathsf{T}}_1,\dots,K^{(i)\mathsf{T}}_C\right]}}{\sqrt{CD/h}}\right).\label{eq:multihead_attention_weights}
\end{align}
$Q_c^{(i)}\in\mathbb{R}^{\frac{d_k}{h}\times T}$, $K_c^{(i)}\in\mathbb{R}^{\frac{d_k}{h}\times T}$, and $V_c^{(i)}\in\mathbb{R}^{\frac{d_v}{h}\times T}$ are calculated from $Q_c$, $K_c$, and $V_c$ using \autoref{eq:q}, \autoref{eq:k}, and \autoref{eq:v}, respectively.
Note that we also omitted the layer index $(n)$ from \autoref{eq:multi_mca}--\autoref{eq:multihead_attention_weights}.

In multi-channel EEND, only the attention calculation in \autoref{eq:multihead_attention_weights} is the inter-channel process, while the other processes are channel-independent.
Also, there are no trainable channel-dependent parameters for the attention calculation in \autoref{eq:multihead_attention_weights}, and the parameters for the channel-independent processes are shared among channels.
Therefore, multi-channel EEND is independent of the number and geometry of microphones.
Note that when the number of channels is one, the multi-head scaled dot-product co-attention in \autoref{eq:multihead_coattention} is identical to the multi-head scaled dot-product attention in \autoref{eq:multihead_attention}, \ie, the co-attention encoder is identical to Transformer encoder.
Moreover, both encoders have the same set of parameters $\vect{\Phi}$ and $\vect{\Psi}$; thus, even if the model is trained as a single-channel model using Transformer encoders, it can handle multi-channel inputs by considering the attention mechanism as a co-attention in \autoref{eq:multihead_coattention}--\autoref{eq:multihead_attention_weights}.

\subsection{Mutual learning of single- and multi-channel EEND}\label{sec:mutual_learning}
In the proposed mutual learning method, given an initial multi-channel model parameterized by $\vect{\Theta}^{(0)}_\text{multi}$, knowledge transfer from multi- to single-channel model and from single- to multi-channel model are iteratively conducted.
More specifically, in the $r$-th round of mutual learning, the following two steps are carried out:
\begin{enumerate}
    \item Train $\vect{\Theta}^{(r)}_\text{single}$ from scratch on $\mathcal{L}_\text{KD}(\vect{\Theta}\mid \vect{X},\vect{\Theta}^{(r-1)}_\text{multi})$,\label{algline:multi_to_single}\\
	\item Train $\vect{\Theta}^{(r)}_\text{multi}$ initialized with $\vect{\Theta}^{(r)}_\text{single}$ on $\mathcal{L}_\text{BCE}(\vect{\Theta}\mid \vect{X},\tilde{Y})$.\label{algline:single_to_multi}
\end{enumerate}
We detail two steps above in \autoref{sec:multi_to_single} and \ref{sec:single_to_multi}, respectively.

\subsubsection{Knowledge transfer from multi- to single-channel EEND}\label{sec:multi_to_single}
To transfer the knowledge from multi-channel EEND to single-channel EEND, we use a knowledge distillation between network outputs (\autoref{fig:overview} left).
According to the observation in \cite{kim2021comparing}, the mean squared error between logits is used as a loss to be minimized instead of Kullback-Leibler divergence.
Given $C$-channel frame-wise acoustic features $\vect{X}=\left[X_c\right]_{c=1}^C$, we calculate the logits of frame- and speaker-wise speech activities $Z_\text{multi}\in\mathbb{R}^{S\times T}$ by using the teacher multi-channel model parameterized by $\vect{\Theta}_\text{multi}^{(r-1)}$ with \autoref{eq:change_dimension}, \autoref{eq:coattentionencoder}, \autoref{eq:average_channel}, \autoref{eq:attractors}, and \autoref{eq:logits}.
We also calculate single-channel results $Z_\text{single}\in\mathbb{R}^{S\times T}$ from randomly selected one of the $C$ channels by using the student single-channel model with \autoref{eq:change_dimension}--\autoref{eq:logits}.
To optimize the student model's parameters $\vect{\Theta}$, here also as in \autoref{eq:bce_loss}, we introduce the following permutation-free knowledge distillation loss to be minimized:
\begin{align}
    \mathcal{L}_{\text{KD}}\left(\vect{\Theta}\relmid \vect{X},\vect{\Theta}_\text{multi}^{(r-1)}\right)=\frac{1}{TS}\min_{P}\norm{Z_\text{multi}-PZ_\text{single}}_F^2,\label{eq:kd_loss}
\end{align}
where $\norm{\cdot}_F$ denotes the Frobenius norm of a matrix and $P\in\left\{0,1\right\}^{S\times S}$ denotes a permutation matrix.
We denote the obtained set of parameters of the student model by $\vect{\Theta}_\text{single}^{(r)}$.

While some studies have investigated the weighted sum of hard-label-based loss $\mathcal{L}_\text{BCE}$ in \autoref{eq:bce_loss} and knowledge-distillation-based loss $\mathcal{L}_\text{KD}$ in \autoref{eq:kd_loss}, we simply used $\mathcal{L}_\text{KD}$ instead of $\mathcal{L}_\text{BCE}$.

\subsubsection{Knowledge transfer from single- to multi-channel EEND}\label{sec:single_to_multi}
As described in \autoref{sec:model}, single- and multi-channel models have the same network parameters.
Therefore, even when a model is trained only on single-channel data, it can handle multi-channel inputs by using co-attention instead of self-attention.
Thus, to transfer the knowledge from a single- to multi-channel model, we simply finetune the single-channel model to a multi-channel model.
We initialize the network parameters with $\vect{\Theta}^{(r)}_\text{single}$ obtained in \autoref{sec:multi_to_single} and finetune the model using the loss $\mathcal{L}_\text{BCE}$ in \autoref{eq:bce_loss} with multi-channel data. The resulted set of parameters of the model is denoted as $\vect{\Theta}^{(r)}_\text{multi}$.

\section{Experiments}
\subsection{Datasets}
To prove the efficiency of the proposed method, we used the simulated two-speaker conversational datasets: SRE+SWBD-train for training and SRE+SWBD-eval for evaluation \cite{horiguchi2022multichannel}, shown in \autoref{tbl:simualted_dataset}.
The simulated conversations were created from single-speaker utterances extracted from 2004--2008 NIST Speaker Recognition Evaluation (SRE) and Switchboard corpora (Switchboard-2 and Switchboard Cellular) (SWBD), noise from MUSAN corpus \cite{snyder2015musan}, and simulated room impulse responses.
For the detailed simulation protocol, refer to \cite{horiguchi2022multichannel}.
While both datasets originally contain 10-channel recordings, we only used one- or four-channel subsets of SRE+SWBD-eval for the evaluation in this paper.

\begin{table}
    \centering
    \setlength{\tabcolsep}{3pt}
    \caption{Two-speaker conversational datasets.}\label{tbl:dataset}
    \vspace{-8pt}
    \subfloat[][Simulated multi-channel datasets.]{\label{tbl:simualted_dataset}%
    \scalebox{\tablescale}{%
    \begin{tabular}{@{}lccccrr@{}}
        \toprule
        &&&\multicolumn{1}{c}{Average}&\multicolumn{1}{c@{}}{Overlap}\\
        Dataset &\#Channels & \#Sessions&\multicolumn{1}{c}{duration}&\multicolumn{1}{c@{}}{ratio}\\\midrule
        SWE+SWBD-train & 10&20,000&\SI{88.7}{\second}&\SI{34.1}{\percent}\\
        SWE+SWBD-eval & 1 / 4&500&\SI{88.1}{\second}&\SI{34.6}{\percent}\\
        \bottomrule
    \end{tabular}%
    }%
    }\\
    \subfloat[][Real single-channel datasets.]{\label{tbl:real_dataset_1ch}%
    \scalebox{\tablescale}{%
    \begin{tabular}{@{}lccrr@{}}
        \toprule
        &&&\multicolumn{1}{c}{Average}&\multicolumn{1}{c@{}}{Overlap}\\
        Dataset &\#Channels & \#Sessions&\multicolumn{1}{c}{duration}&\multicolumn{1}{c@{}}{ratio}\\\midrule
        CH-2spk Part 1&1 &155& \SI{74.0}{\second}&\SI{14.0}{\percent}\\
        CH-2spk Part 2&1 &148&\SI{72.1}{\second}& \SI{13.1}{\percent}\\
        CSJ &1&54&\SI{766.6}{\second}&\SI{20.1}{\percent}\\
        \bottomrule
    \end{tabular}%
    }%
    }\\
    \subfloat[][Real multi-channel datasets.]{\label{tbl:real_dataset_4ch}%
    \scalebox{\tablescale}{%
    \begin{tabular}{@{}lccrr@{}}
        \toprule
        &&&\multicolumn{1}{c}{Average}&\multicolumn{1}{c@{}}{Overlap}\\
        Dataset &\#Channels & \#Sessions&\multicolumn{1}{c}{duration}&\multicolumn{1}{c@{}}{ratio}\\\midrule
        CSJ-train & 9& 100&\SI{113.5}{\second}&\SI{11.0}{\percent}\\
        CSJ-eval & 4 & 100&\SI{102.2}{\second}&\SI{9.6}{\percent}\\
        CSJ-dialog & 4& 58&\SI{755.2}{\second}&\SI{17.3}{\percent}\\
        \bottomrule
    \end{tabular}%
    }%
    }
\end{table}

With domain adaptation on real datasets, we show that models trained on the simulated dataset using the proposed method are also good pretrained models.
As the single-channel datasets, we used the two-speaker subset of CALLHOME (CH-2spk) and the dialog part of the corpus of spontaneous Japanese (CSJ) \cite{maekawa2003corpus} following \cite{horiguchi2022encoderdecoder}, as shown in \autoref{tbl:real_dataset_1ch}.
CH-2spk was split into two subsets (Part 1 \& 2) according to the Kaldi CALLHOME diarization recipe\footnote{\url{https://github.com/kaldi-asr/kaldi/tree/master/egs/callhome_diarization/v2}}.
CH-2spk Part 1 was used as the adaptation set and CH-2spk Part 2 and CSJ were used as the evaluation sets.
We regard CH-2spk Part 2 as an in-domain dataset and CSJ as an out-of-domain dataset.

As the multi-channel datasets, we used  CSJ-train, CSJ-eval, and CSJ-dialog\footnote{CSJ in \autoref{tbl:real_dataset_1ch} and CSJ-dialog in \autoref{tbl:real_dataset_4ch} share the data source but the numbers of sessions are slightly different. This is because four of 58 sessions were eliminated in the single-channel CSJ following the conventional study \cite{horiguchi2022encoderdecoder}.} \cite{horiguchi2022multichannel}, each of which is shown in \autoref{tbl:real_dataset_4ch}.
These are re-recorded datasets in which each speaker's utterances were played via loudspeakers placed in a meeting room and recorded with distributed smartphone devices.
CSJ-train was used as the adaptation set and CSJ-eval and CSJ-dialog were used as the evaluation sets.
Since CSJ-train and CSJ-eval are simulated conversations created using single-speaker recordings while CSJ-dialog is actual conversations, we regard CSJ-eval as an in-domain dataset and CSJ-dialog as an out-of-domain dataset.
Note that the original CSJ-eval and CSJ-dialog have nine channels, but we only used four-channel subsets for our experiments.

We conducted our experiments using two-speaker datasets because this paper is aiming at proving the effectiveness of the proposed mutual learning method, but we note that EEND-EDA can handle the case where the number of speakers is unknown.

\subsection{Settings}
The inputs to single- and multi-channel models are 345-dimensional features extracted every \SI{100}{\ms} for each channel, which are prepared with the following procedure:
\begin{enumerate}
    \item Extract 23-dimensional log-mel filterbanks for each \SI{10}{\ms},
    \item Apply frame splicing ($\pm7$ frames) for them, resulting in 345 dimensions,
    \item Subsample them by a factor of 10.
\end{enumerate}

We used a four-layer encoder (\ie, $N=4$ in \autoref{sec:overview}) for each EEND model with the dimensionality of intermediate embeddings $D=256$, and the number of heads in each encoder $h=4$.
The baseline single- and multi-channel EENDs were trained from scratch using SRE+SWBD-train.
During training, one or four of ten channels were randomly selected and fed to the models, respectively.
Each model was optimized for 500 epochs with Adam \cite{kingma2015adam} with the Noam scheduler \cite{vaswani2017attention} with warm-up steps of 100,000.
For the training of the baseline multi-channel EEND, we used the channel dropout technique \cite{horiguchi2022multichannel}.

For the knowledge distillation from multi-channel to single-channel models, four channels of SRE+SWBD-train out of ten were selected and fed to the multi-channel model to obtain $Z_\text{multi}$ in \autoref{eq:kd_loss}, and one of the four channels was input to the single-channel model to obtain $Z_\text{single}$ in \autoref{eq:kd_loss}.
The same training strategy used to train the baseline models was used for knowledge distillation.
Note that the single-channel model here was trained from scratch.

For the finetuning of the single-channel model using multi-channel data, four of ten channels of SRE+SWBD-train were randomly selected at each iteration and fed to the model.
The model was finetuned for another 100 epochs with Adam using the Noam scheduler with 20,000 warm-up steps.

For the adaptation on the real-recorded datasets, we used CH-2spk Part 1 for single-channel evaluation or CSJ-train for multi-channel evaluation, respectively.
Adam optimizer with a fixed learning rate of $1\times 10^{-5}$ was used for 100 epochs of adaptation.

We report diarization error rates (DERs) with \SI{0.25}{\second} collar tolerance.
Note that speaker overlaps are included in the evaluation.
While conventional studies generally apply a median filter as post-processing, in order to clarify the extent to which frame-level accuracy has been improved, this paper discusses the results without median filtering.
For reference, the results with 11-frame median filtering are also shown in parentheses in each table.

\subsection{Results}
\subsubsection{Results on the simulated dataset}

\begin{table}[t]
    \centering
    \caption{DERs (\%) improvement on SWE+SWBD-eval with the proposed method. The values in gray indicate the mismatched condition in the number of channels. The values in the parentheses are with median filtering.}
    \label{tbl:results_simu}
    \scalebox{\tablescale}{%
    \begin{tabular}{@{}lc@{\hskip3pt}cc@{\hskip3pt}c@{}}
        \toprule
        Method&\multicolumn{2}{c}{\dersingle}&\multicolumn{2}{c@{}}{\dermulti}\\\midrule
        \ctext{1} Baseline 4-ch model&\gray{5.79} &\gray{(4.96)}&2.32 &(1.98)\\
        \ctext{2} Baseline 1-ch model&4.11& (3.91)&\gray{4.90} &\gray{(4.44)}\\
        \ctext{3} Finetune \ctext{2} using 4-ch data&\gray{12.76} &\gray{(11.16)}& 2.67 &(2.49)\\\midrule
        \ctext{4} Knowledge distillation from \ctext{1}&  3.34 &(3.17)&\gray{4.04} &\gray{(4.40)}\\
        \ctext{5} Finetune \ctext{4} using 4-ch data& \gray{12.11} &\gray{(9.83)}&2.17 &(2.02)\\\midrule
        \ctext{6} Knowledge distillation from \ctext{5}&3.08 &(2.94)&\gray{3.57} &\gray{(3.25)}\\
        \ctext{7} Finetune \ctext{6} using 4-ch data&\gray{10.83} &\gray{(8.73)}&2.08 &(1.94)\\
        \bottomrule
    \end{tabular}%
    }
\end{table}

\autoref{tbl:results_simu} shows the DER improvement using the proposed mutual learning on SWE+SWBD-eval data.
The DERs under mismatched conditions, \ie, single-channel (four-channel) results using the model trained on the four-channel (single-channel) data, are written in gray.
For clarification, we denote the DERs evaluated using single-channel SWE+SWBD-eval as \dersingle and those evaluated using four-channel data as \dermulti.

The first and the second rows show the DERs of the baseline single-channel and four-channel models, respectively.
It is clearly observed that the models can still decode data of mismatched conditions, but the resulting DERs are worse than those of matched conditions.
Finetuning of the baseline single-channel model on four-channel data improved the \dermulti from \SI{4.11}{\percent} to \SI{2.67}{\percent} as in the third row, but it did not reach the \dermulti of the baseline four-channel model (\SI{2.32}{\percent}).
These results indicate that single-channel pretraining does not always improve the performance of the multi-channel model.
We decided to use the baseline four-channel model \ctext{1} as the initial model of mutual learning.

When we trained a single-channel model with the baseline four-channel model as a teacher using knowledge distillation, the \dersingle was improved from \SI{4.11}{\percent} to \SI{3.34}{\percent} as in the fourth row.
Finetuning the model \ctext{4} with four-channel data resulted in the \dermulti of \SI{2.17}{\percent}, outperforming the baseline four-channel model \ctext{1}.
This bi-directional improvement between single- and multi-channel models has proven the effectiveness of the proposed mutual learning.

We then ran another round of mutual learning starting from the model \ctext{5}.
Since the improvement of \dermulti from \ctext{1} to \ctext{5} was $0.15$ ($=2.32-2.17$) percentage points, the improvement of \dersingle between their distilled models was larger: $0.26$ ($=3.34-3.08$).
Finetuning the distilled model \ctext{6} with four-channel data led to a further improvement in \dermulti, which was \SI{2.08}{\percent} as in the seventh row.

\subsubsection{Results on the real datasets}
\begin{table}[t]
    \centering
    \caption{DERs (\%) on single-channel real conversational datasets.}
    \label{tbl:results_real_1ch}
    \scalebox{\tablescale}{%
    \begin{tabular}{@{}cc@{\hskip3pt}cc@{\hskip3pt}c@{}}
        \toprule
        &\multicolumn{4}{c}{Evaluation dataset}\\\cmidrule(l){2-5}
        Pretrained model&\multicolumn{2}{c}{CH-2spk Part 2}&\multicolumn{2}{c}{CSJ}\\\midrule
        \ctext{2}&14.11& (12.56)& 24.71&(24.15)\\
        \ctext{4}&10.06& (8.95)& 22.52&(22.09)\\
        \ctext{6}&9.85& (8.45)& 21.33 &(20.66)\\
        \bottomrule
    \end{tabular}%
    }
\end{table}

\begin{table}[t]
    \centering
    \caption{DERs (\%) on multi-channel real conversational datasets.}
    \label{tbl:results_real_4ch}
    \scalebox{\tablescale}{%
    \begin{tabular}{@{}cc@{\hskip3pt}cc@{\hskip3pt}c@{}}
        \toprule
        &\multicolumn{4}{c}{Evaluation dataset}\\\cmidrule(l){2-5}
        Pretrained model&\multicolumn{2}{c}{CSJ-eval}&\multicolumn{2}{c@{}}{CSJ-dialog}\\\midrule
        \ctext{1}&1.19& (0.66)&16.14&(15.78)\\
        \ctext{3}&1.51& (0.93)&18.77&(18.07)\\
        \ctext{5}&1.17& (0.67)&17.30&(16.60)\\
        \ctext{7}&1.08& (0.66)&16.98&(16.42)\\
        \bottomrule
    \end{tabular}%
    }
\end{table}

We then evaluated each single-channel model, \ie, \ctext{2}, \ctext{4}, and \ctext{6} in \autoref{tbl:results_simu}, with the adaptation on CH-2spk Part 1.
The DERs on the CH-2spk Part 2 and CSJ datasets are shown in \autoref{tbl:results_real_1ch}.
It is clearly observed that pretraining using the proposed mutual learning method also improved the performance on the real datasets, regardless of in-domain or out-of-domain.

Each multi-channel model, \ie, \ctext{1}, \ctext{3}, \ctext{5}, and \ctext{7} in \autoref{tbl:results_simu}, was also adapted using CSJ-train.
The results on CSJ-eval and CSJ-dialog using those adapted models are shown in \autoref{tbl:results_real_4ch}.
For CSJ-eval, the pretrained models trained using the proposed method, \ie, \ctext{3}, \ctext{5}, and \ctext{7}, helped to improve the DERs gradually with the best DER of \SI{1.08}{\percent}, which also outperformed the model based on \ctext{1}.
In the case of CSJ-dialog, the DERs were also reduced with respect to the pretrained model finetuned from the single-channel model, \ie, \ctext{3}, \ctext{5}, and \ctext{7}.
However, the most accurate model was based on \ctext{1} in terms of CSJ-dialog.
Since CSJ-dialog is the out-of-domain dataset, it is more advantageous to rely on spatial information.
That is why the model based on \ctext{1} performed best on CSJ-dialog because \ctext{1} was trained using multi-channel data from the beginning and is considered to be highly dependent on spatial information.

\section{Conclusion}
In this paper, we proposed a mutual learning method of single- and multi-channel models.
With the model that can treat both single- and multi-channel inputs, we alternately execute 1) knowledge distillation from a multi-channel model to a single-channel model and 2) finetuning from the distilled single-channel model to a multi-channel model.
Experimental results showed that the proposed method gradually improved DERs on the single- and multi-channel conditions.
Future work will apply this method to other speech processing tasks such as speech separation and recognition.

\bibliographystyle{IEEEbib-abbrev}
\bibliography{mybib}

\end{document}